\documentclass{aa}  
\usepackage{natbib}
\bibpunct{(}{)}{;}{a}{}{,} 
\usepackage{longtable}

\usepackage{graphicx}

\usepackage{txfonts}
\usepackage{fixltx2e}

\begin{document}

   \title{Molecular transitions as probes of the physical conditions of extragalactic environments}


   \author{Serena Viti
          \inst{1}
          }

   \institute{Department of Physics and Astronomy, University College London,
              Gower street, London, WC1E 6BT, UK\\
              \email{serena.viti@ucl.ac.uk}
             }

   \date{Received; accepted}

 
  \abstract
   {}
   {We present a method to interpret molecular observations and molecular line ratios in nearby extragalactic regions. }
   {Ab initio grids of time dependent chemical models, varying in gas density, temperature, cosmic ray ionization rate, and radiation field, are used as inputs into RADEX calculations. Tables of abundances, column densities, theoretical line intensities, and line ratios for some of the most used dense gas tracers are provided. The degree of correlation as well as degeneracy inherent in molecular ratios is discussed. Comparisons of the theoretical intensities with example observations are also provided. }
   {We find that, within the parameters space explored, chemical abundances can be constrained by a well-defined set of gas density, gas temperature, and cosmic ray ionization rates for the species we investigate here.  However, line intensities, and more importantly line ratios, from different chemical models can be very similar, thereby leading to a clear degeneracy. We also find that the gas subjected to a galactic cosmic ray ionization rate will not necessarily have reached steady state in 1 million years. The species most affected by time dependency effects are HCN and CS, which are both high density tracers. We use our ab initio method to fit an example set of data from two galaxies, i.e. M82 and NGC~253. We find that (i) molecular line ratios can be easily matched even with erroneous individual line intensities, (ii) no set of species can be matched by a one-component interstellar medium (ISM), and (iii) a species may be a good tracer of an energetic process but only 
under specific density and temperature conditions.}
   {We provide tables of chemical abundances and line intensities ratios for some of the most commonly observed extragalactic tracers of dense gas for a grid of models. We show that by taking the chemistry behind each species and the individual line intensities into consideration, many degeneracies that arise by just using molecular line ratios can be avoided. Finally we show that using a species or a ratio as a tracer of an individual energetic process, such as cosmic rays and UV, ought to be done with caution. }


   \maketitle
%

\section{Introduction}

Following the first detection of a molecular species in external galaxies in the 1970s, 
molecular line emission studies have been routinely performed to determine the physical characteristics of nearby galaxies \citep[e.g.][]{Takano02,Mart09,Alad11,Alad13}. 
Galaxies span a variety of shapes, sizes,
and physical conditions. The physical parameters, such as gas densities,
ultraviolet (UV) fields, cosmic ray ionization rates, and dust properties, which determine the appearance of our own Galaxy may therefore be very different in other galaxies. 
Nevertheless external galaxies, in
general, contain the variety of regions and sources similar to those
that we can identify in the Milky Way; hence
molecules could be ideal probes of the dense gas that is
the reservoir for star formation and could help in describing the process of star formation
itself as well as the influence of newly formed stars on their
environments \citep{WV14}.

Because of their range of critical densities across the different molecular species and across the transitions
of the same molecule, molecular observations are an ideal tool to trace a wide range of gas densities
in the interstellar medium and the dependencies of chemical reactions on the energy available to the system, i.e the kinetic temperature of the gas.
However, what makes observations of molecules so adept is also what complicates their interpretation. In particular, even with the highest spatial resolution available with the Atacama Large Millimeter/Submillimeter Array (ALMA), the beam usually encompasses emissions from many types of regions
and the molecular emission is an ensemble of multi-phase gas in which the spatial
and temporal effects are diluted in the beam. Concomitant detection of multiple molecules in a spatially unresolved
galaxy therefore does not necessarily mean that they are emitted from the same gas \citep[e.g.][]{Viti15}.
While this is not surprising, many multi-species studies of nearby galaxies still need to assume coexistence of molecular emission when
interpreting their observations \citep[e.g.][]{User04,Krip11,Viti14}.
This is partly because we are forced to use multiple transitions and molecules to overcome the well-known degeneracy
between density and temperature for linear molecules, which implies that single transitions of these molecules
cannot be easily used to determine quantitatively the gas density or temperature. 
There have been several theoretical studies on the sensitivity of chemistry to the range of physical parameters that determine a galaxy \citep[e.g.][]{Baye08,Baye09,Baye11,Meij11,Kaza12,Kaza15}. Most of these sensitivity studies concentrated on the modelling of photon-dominated regions (PDR) or X-ray dominated regions (XDR), rather than the modelling of dense molecular gas.
Observationally, many studies have shown that molecular line ratios, such as HCO$^{+}$/HCN or HCN/CO,
differ across different types of galaxies especially between Active Galactic Nuclei (AGN)-dominated galaxies and starburst-dominated galaxies 
\citep[e.g.][]{Kohno05,Krip08},
although some studies have found that enhanced HCN/HCO$^+$ are not unique to AGN environments but can also be found in systems dominated by star formation \citep[e.g.][]{Priv15}. However the derived abundance ratios for an individual galaxy also highly differ across studies depending on the transitions observed, available resolution, and method used in deriving the column densities.
An understanding of the chemistry behind each molecule and its dependencies on the density and temperature of the gas is therefore essential and will lead to a more coordinated
approach to understanding the nature of galaxies using molecular line
emissions.

In this paper we use, as a test bed, a relatively small grid of chemical models to present a method for the
interpretation of molecular observations and molecular line ratios in nearby extragalactic regions, in particular, starburst and AGN dominated galaxies as these are the most targeted types of galaxies for dense gas studies; similar studies can be repeated for Ultra Luminous Infrared Galaxies (ULIRG), dwarf galaxies, etc., where, however, much of the gas may be in ionic or atomic form \citep[e.g.][]{vander10,WV14}. We analyse the theoretical abundances and column densities of some of the most common tracers for a parameter space in gas density, temperature, cosmic ray ionization rate, and radiation field.  The theoretical abundances are then used as inputs to the radiative transfer code RADEX \citep{VanderTak07}, which in turn provides the theoretical intensities as well as common line ratios for the parameter space investigated by the chemical models. We investigate the degree of correlation and degeneracy inherent in each ratio and show how one can use theoretically derived abundances and line intensities for comparison with observations by using published observational data as examples.  

The aim of this study is twofold. 
First, we seek to aid the observer in determining the most likely scenario that fits the data when multiple solutions are present. Second, we aim to provide online tables that one can use directly to compare with their observed intensities and intensity ratios for the most common used transitions; the tables are simply meant to be an example subset but include some of the most observed molecular species and transitions. We present the chemical models and discuss their degeneracies in Section 2. In Section 3 we first calculate theoretical line intensities and line ratios for four of the most observed gas tracers: CO, HCN, HCO$^+$, and CS. While the focus of this paper is purely on the characterization of the dense gas, CO (whose critical density is low and hence ubiquitously traces all gas above $\sim$ 100 cm$^{-3}$) is included because it is still routinely used to interpret molecular ratios or determine the star formation rate density \citep[e.g.][]{Kohn01,Kohno05,Tan2013}, especially if high-J CO observations are available \citep{Zhao2016}. Secondly, for completeness, we also extend our abundance and line intensities analyses to further species, not as commonly observed in external galaxies as those above, but often used as tracers of particular energetic events (e.g. shocks). In Section 4 we make use of our calculations to partly reinterpret observations in two very well studied galaxies to give examples of how this methodology can be used.
\par
All the tables referred to in the paper are online and can be found at https://uclchem.github.io/viti2017.pdf.


\section{Chemical models: Theoretical molecular abundances}
We used the {\tt UCL\_CHEM\footnote{This code has now being made public \citep{Hold17} and can be downloaded at https://uclchem.github.io/}} gas-grain time dependent chemical code \citep{Viti04} to run a grid of chemical models covering a range of gas densities (10$^4$--10$^6$ cm$^{-3}$), temperatures (50--200 K), radiation fields (1--500 Draine), cosmic ray ionization rates (1--500 standard galactic cosmic ray ionization field), and visual extinctions (1--50 mags); we used the cosmic ray ionization flux to simulate an enhancement in X-ray flux. 
As previously noted \citep{Viti14}, this approximation has its limitations in that the X-ray flux heats the gas more efficiently than cosmic rays. However, the chemistry arising from these two fluxes should be similar. 

While there are several more parameters that can be explored when performing chemical modelling, for the purpose of this study we chose those that affect most the line intensities of dense gas tracers in starburst and AGN-dominated galaxies, as predicted by Large Velocity Gradient (LVG) models. For example, gas density, temperature, and visual extinction (directly correlated to column density) are all input parameters to RADEX and other LVG models. Cosmic ray ionization and radiation fluxes are known to differ from the Galactic values in starburst and AGN-dominated galaxies, and hence are also considered. Moreover they directly affect the temperature of the gas and dust. The choice of the lower and upper limits for the gas densities and temperatures reflect the best fits found for the dense molecular gas in the most commonly studied starburst and AGN galaxies; they are by no means meant to represent the only possible values. The range of cosmic ray ionization rates and radiation fields are subsets of the wider range explored in previous theoretical studies. As already mentioned \citep[e.g.][]{Baye09, Baye11, Meij13}, our grid of models is not meant to be exhaustive but simply represent a plausible range of physical conditions.   We also ran some models simulating the presence of shocks. For these models we adopted a representative shock velocity of 40 km s$^{-1}$, which depending on the pre-shock densities, leads to a maximum temperature for the shock ranging from 2200 K to 4000 K. The shocks in UCL\_CHEM are treated in a parametric form as in \citet{Jim08}. Below we briefly describe UCL\_CHEM. Further details of this version of UCL\_CHEM, coupled with the shock module, can be found in \citet{Viti11}, and \citet{Hold17}.

UCL\_CHEM is an ab initio model and computes the evolution as a function of time of chemical abundances of
the gas and on the ices starting from a diffuse and atomic gas (at a density of 10 cm$^{-3}$). The parameter regulating the visual extinction has two effects on the models. First, 
this parameter affects the photochemistry since the radiation field is very efficiently attenuated for A$_V$ $>$ 4-5 mags.  However, and more importantly, this paramater
is effectively a measure of the size of the emission region, since in order to derive column densities from the fractional abundances calculated by the models we multiply the latter by the hydrogen column density as measured at 1 magnitude times the visual extinction. This is the so-called “on the spot” approximation
\cite[e.g.][]{dyswill97}. As the chemical model is single point, we estimate the theoretical column densities, a posteriori. 

The code is run in two phases in time (Phase I and Phase II). The initial density in Phase I is $\sim$ 10 cm$^{-3}$. The gas is allowed to collapse and to reach a high density gas by means of a free-fall collapse.
The collapse is not meant to represent the formation of protostars, but it is simply a way to compute the chemistry of high density gas in a self-consistent way starting from a diffuse atomic gas, i.e. without assuming the chemical composition at its final density. Of course, it is unknown how the gas reached its high density and we could have assumed any other number of density functions with time. For simplicity, we adopt free-fall collapse.  The temperature during this phase is kept constant at 10 K, and the cosmic ray ionization rates and radiation fields are at their standard Galactic values of $\zeta_o$ = 5$\times$10$^{-17}$ s$^{-1}$ and 1 Draine, or 2.74$\times$10$^{-3}$ erg/s/cm$^{2}$, respectively. In this way the chemistry of the high density gas is computed in a self-consistent way starting from a diffuse atomic gas, i.e without assuming the chemical composition at its final density. During Phase I 
atoms and molecules are allowed to freeze onto the dust grains and react with each other, forming icy mantles. In Phase II, for the non-shock models, 
UCL\_CHEM computes the chemical evolution of the gas and the icy mantles after either an assumed burst of star formation or AGN activity has occurred. The temperature of the gas increases from 10~K  to a value set by the user and sublimation from the icy mantles occurs. The chemical evolution of the gas is then followed for 10$^7$ years. In the shock models, Phase II considers a plane-parallel C-Shock propagating with a velocity, $V_s$, through the ambient medium.  Details of the code and its use for the determination of abundances in external galaxies can be found in \citet{Viti14}. A detailed description of the shock module in UCL\_CHEM can be found in \citet{Viti11} and in \citet{Jim08}.
In both phases of the UCL\_CHEM model, the gas phase chemical network is based on the UMIST13 database \citep{McElroy2013} with updates from the KIDA database \citep{Wakelam2015}. The surface reactions included in this model are assumed to be mainly hydrogenation reactions, allowing chemical saturation when possible.

Tables 1 and 2 list the grid of models with  the resultant abundances (Table 1), column densities, and ratios (Table 2) for four of the most observed molecules in galaxies: CO, HCO$^+$, HCN, and CS. 
Both tables show the results at the final time step (10$^7$ years). However for the shock models, the shock occurs at the beginning of Phase II and the models are stopped once the dissipation length has been reached; hence the Tables show the chemistry at the last time step, representing the post-shocked gas (at $\sim$ 10$^5$ years). For the shock models the $A_V$ is the initial visual extinction before the passage of the shock. In Table 3 we tabulate the column densities for selected models at an earlier time of 1 Myr (see Section 2.2). 

Tables 4, 5, and 6 list the column densities for further species, 
often used as tracers of shocks (e.g. SiO, CH$_3$OH, and SO) or surface grain chemistry (e.g. HNCO). HNC has been added because several studies use the HNC/HNC ratio as a tracer of dense PDR and XDRs environments \citep[e.g.][]{Green16}. We note that HNCO and CH$_3$OH do not
survive until chemical equilibrium is reached and they are only listed for 10$^6$ years for all non-shock models. Table 6 lists the column densities for CH$_3$OH and HNCO for Models 65-67 at two different times: during the peak temperature of the shock and once the shock has passed. For the latter time the column densities were calculated assuming an average visual extinction of 10 mags.

\subsection{Chemical 'degeneracy'}
From Tables 1--6 we find a chemical degeneracy worth mentioning in relation to this study.  
Models differing by one or more parameters can still lead to column densities similar 
enough to be within the observational error in the column density estimation. For example, the HCN column density remains the same when cosmic ray ionization rate is enhanced from 1 to 10 $\times$ the Galactic value (cf. Models 1 and 9); however this does not mean HCN is not sensitive to changes in the cosmic ray ionization rates as higher cosmic rate affect the HCN chemistry. In most 
cases this degeneracy can be dealt with if one
observes multiple molecular species at the same spatial resolution. For example, Models 3 and 6 lead to almost identical CO,  and similar HCO$^+$ and HCN column densities but they differ by $\sim$ one order of magnitude in CS.   It is key to know a priori which species one needs to observe to differentiate among models by referring to Tables such as those presented here.  
However a more interesting degeneracy is caused when one 
obtains the same line intensities for different combinations of [n$_H$,T$_{gas}$, $N(X)$]. We discuss this in Section 3.

\subsection{The importance of time dependency}

One chemical effect that is difficult to discern observationally is the time dependency of the abundance of each species. We do not discuss the shock models in this section, as chemical equilibrium is never reached during the passage of the shock. 
As shown by several studies in the past, not all chemistry is equally affected by time. Hence, while for some of the models we ran, equilibrium is reached well before 1 million years, for others it is not reached until much later; this is important because it is not 
obvious whether the large scale gas we observe, especially in starburst galaxies, will have remained subjected to the same energetic and dynamical conditions for longer than 1 million years because, for example, the flux of cosmic ray will vary across episodes of starbursts. Time dependencies effects have been discussed before \citep{Baye08,Meij13}, in the context of specific galaxies as well \citep[e.g.][]{Alad13}. In Tables 3 and 5  we tabulate the column densities for the models where chemical equilibrium has not been reached yet at 1 million years, and hence where the column densities differ from Tables 2 and 4.  
In Tables 3 and 5 there are no models with $\zeta$ higher than the standard value;  
it is clear therefore that a high cosmic ray ionization rate leads to steady state more quickly, as also found by previous work \citep{Baye11}. On 
the other hand, an enhanced radiation field (at least up to 500 the interstellar galactic field) does not help the chemistry in reaching steady state at the visual extinctions considered here, since the UV does not penetrate. 
Interestingly, of the four most observed species (CO, HCN, HCO$^+$, and CS), the species most affected by time dependency is HCN followed by CS: both of these species are routinely used as high density gas tracers. 

Some species are always short-lived; in the parameter space explored,  these are CH$_3$OH and HNCO. 
In all the non-shock models these two species are always destroyed by the time chemical equilibrium is reached. In models not listed in Table 5, these species do not even survive for a 1 Myr; hence an observed high abundance of these two species is either a signature of a young gas or, more likely, a dynamically excited gas. In fact, we find that for some of the models in which HNCO and/or CH$_3$OH are very low before or at  10$^6$ years, their abundance is well enhanced at earlier times.
\section{Radiative transfer calculations: Intensities and intensities ratios}

We estimated the theoretical integrated intensity for the first 10 rotational lines of CO, HCN, HCO$^+$, and CS, and for selected lines of SiO, HNC, SO, CH$_3$OH, and HNCO, where the line selection is observationally motivated. 
The intensities were calculated using the radiative transfer code RADEX \citep{VanderTak07} from the grid of models in Tables 2-6. The collisional rates used for the RADEX calculations were taken from the LAMBDA database \citep{Scho05} and were computed with H$_2$ as collisional partner, where available. We used a background temperature of 2.7~K. The approach we take in this paper differs from the usual use of RADEX or any LVG modelling in external galaxies in that we do not produce a grid in order to fit a set of observations. Rather, we used the input from our chemical modelling to fix the three-parameter set [n$_H$, T$_{gas}$, $N(X)$]. Hence the only two free parameters in RADEX are the linewidth and geometry. We take two representative linewidths of 50 and 100 kms$^{-1}$, but as the results do not affect our main conclusions we only present the 100 kms$^{-1}$ models. Although RADEX can be used with different geometries, for the purpose of these calculations we used 
spherical geometry. While with Tables 2-6 such calculations can be repeated for more geometries, \citet{Krip11} used
all three geometries available to characterize the dense gas in the AGN dominated nucleus of NGC~1068 but did not find significant differences in their fitting.

Also, estimating theoretical 
intensities for the shock models using RADEX falls well short of a proper radiative transfer calculation since the gas temperatures and densities vary quickly with the passage of the shock, and so do the abundances; hence one must be aware that these line intensities represent a gas that has been shocked (by a single or multiple shock episodes) but that has cooled back down to its 'equilibrium' temperature (in this case 100~K). Clearly a large variation of the parameter space can be performed when modelling shocked gas, which goes beyond the scope of this paper.   

The results of our calculations are tabulated in Tables 7 and  8, for the most commonly observed species, namely CO, HCO$^+$, HCN and CS; in Table 7 we list chemical abundances
as obtained at 10$^7$ years or at the last time step in the case of the shock
models and in Table 8 we list chemical abundances as obtained at 10$^6$ years for the models in which
chemical equilibrium was not reached at this time. In Tables 10--16 we tabulate selected line intensities for SiO, SO, HNC, and HNCO. Because of the large number of CH$_3$OH transitions observed as $k$ bands, we chose not to tabulate its line intensities (which can however be easily derived from the column densities tables.)   For some models, at 10$^6$ years, the RADEX calculation of the HCN does not converge; what these models have in common, at 1 million years, is that the HCN column density is very high and the gas is either very dense (10$^6$ cm$^{-3}$) and/or very hot (200 K).  The issue of non-convergence is most likely due to high-opacity effects, and in these cases Monte Carlo techniques combined with 
an accelerated lambda
convergence method have been shown to perform better \citep[e.g.][]{Mills2013}.

What it is often used in molecular studies in external galaxies are molecular line ratios. In Table 9 we tabulate the HCN/HCO$^+$ ratios for each transition; we deliberately approximate to its integer value as it is seldom that one can claim a better accuracy from observational values. It is clear even from a quick glance that the often used ratios, for example the HCN(4-3)/HCO$^+$ (4-3) or the HCN(1-0)/HCO$^+$(1-0), are the same for several models within the
accuracy assumed. While this is unavoidable, by approaching the interpretation of the observations  with an a priori method such as presented here, one may, at least, be able to constrain this degeneracy. As a practical example, let us consider the HCN/HCO$^+$ line ratios, which we plot in Figure 1 for four different transitions. These ratios are routinely used as tracers of AGNs because they are found to be significantly higher in AGN when compared to those found in starbursts \citep[e.g.][]{Krip08}. 
Model 1 has the same HCN/HCO$^+$(1-0) and similar HCN/HCO$^+$(2-1) ratio to Model 7 but all the other ratios are very different. These two models differ substantially only in gas density, which is a physical characteristic that
the lower transition ratios could not have traced.  
Figure 1 shows  how a relatively standard grid of chemical models can lead to large ranges for the HCN(1-0)/HCO$^+$ ratio of key transitions.  This large range of ratios is consistent with the fact that, as pointed out in \citet{Krip08}, among others, differences in the intensity ratios may arise due to different densities, temperatures, radiation fields, and evolutionary state of activity. However, as long as more than two transitions per molecules are observed, degeneracies can indeed be broken, at least within the parameter space explored here.

\begin{figure*}
\includegraphics[angle=-90,scale=0.7]{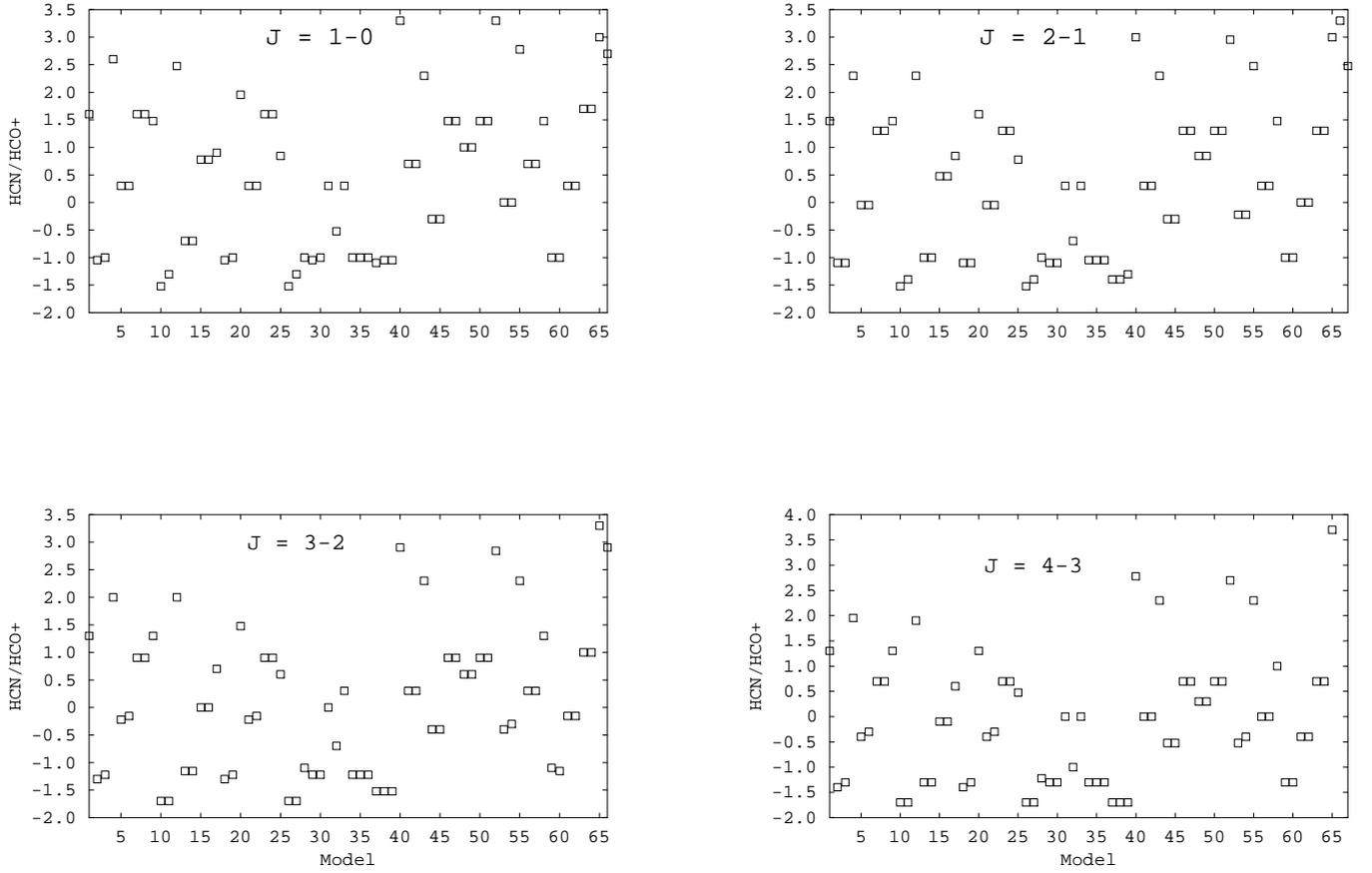}
\caption{Theoretical log line ratio HCN/HCO$^+$ up to J = 4 - 3 for each of the chemical model. }
\end{figure*} 

The approach we outline here is not proposed as a substitute for the more canonical use of LVG modelling but as a complementary analysis. In particular one could use tables, such as those presented in this study, to determine a priori whether similar line intensities arise from very different physical conditions and viceversa. This may be of particular importance in light of the results from \citet{TG16} who find that LVG models struggle to recover any parameter better than to within half a dex and they find no evidence of systemic offsets. 
While these Tables are not exhaustive we hope that they can be used to track the origin of multiple solutions.

\section{Characterizing the dense gas in nearby galaxies}
In this section we provide examples of how one can use theoretically determined molecular intensities and ratios to interpret observations.  We refer to two commonly observed nearby galaxies as our test beds and we compare an arbitrary selection of single-dish observations for one of the galaxies, and interferometric observations for the second galaxy. What is presented here is not a full model of any of these two galaxies. In order to do that a much more 
comprehensive set of observations, together with a larger grid of models, ought to be used.  
\subsection{M82}
M82 (at a distance of $\sim$ 3.50 Mpc) is one of the most studied nearby starburst galaxies. Among the many works, we chose, as an example, the observations presented in \citet{Alad11} who performed a line survey of the north-east (NE) lobe of M82 with the single dish IRAM 30 m telescope. The spatial resolution obtained ranged from 158 to 333 kpc.  

Of the molecular transitions they observed, we first compared CS (3-2), CS (5-4), and the HCO$^+$(3-2). We directly compared the integrated intensities (Table A.1 in \citet{Alad11} by making the crude assumption that the emission region fills the beam. In fact, according to \citet{Alad11}, and based on previous interferometric observations, the emission region should be within a factor of 1--1.5 of the beam size, hence our assumption is justified. However, substructures likely still exist within the beam and hence the observed line temperatures may be in fact smaller than the actual temperatures. 
The CS (3-2), observed to have an integrated line intensity of $\sim$ 11.5 K km s$^{-1}$, 
is reasonably well matched by Models 8, 24, and to a lesser extent, 32. The first two models simulate very dense gas (10$^6$ cm$^{-3}$) and only differ in the strength of the radiation field, while the latter simulates a lower density but cosmic ray dominated gas. From this transition and single dish observations alone, we cannot distinguish between the two types of gas. On the other hand, for the CS (5-4) transition with an observed integrated line intensity of $\sim$ 4.8 K km s$^{-1}$, the closest (but not a good fit) match is achieved by Models 49, 7 and 23; these all have a very high density gas but differ from those matching the lower $J$ transition either in visual extinction or temperature. 
The HCO$^+$ (3-2) transition, observed to have an intensity of $\sim$ 4.2 K kms$^{-1}$, is instead 
best matched by Model 44, simulating a low density (10$^4$ cm$^{-3}$), high temperature (200 K), and  a low visual extinction gas, as one would expect as ions do not survive in very optically thick gas unless the ionization rates are enhanced. In fact, Model 44 best matches the HCO$^+$ (3-2) transition at earlier times. 

Of course, molecular ratios of lines are routinely used. This however makes the assumption that the two transitions arise from the same gas and that the individual 
column densities are both correct. For example, the CS(5-4)/CS(3-2) ratio is best matched by Models 13, and then 6 and 22; none of these models match either of the individual line intensities. 

Finally, comparing the CS LTE column density derived by \citet{Alad11} with our best matching models we find that all models matching the CS (3-2) can reproduce it, but, of the models that best match the CS (5-4) transition, Model 49, with the highest temperature and visual extinction, matches best the column density. The column density of HCO$^+$ is very well matched by the model that fits best its line intensity.  

\citet{Alad11} also observed SO(4$_3$-3$_2$) and several methanol transitions. For the former its line intensity (0.94 K km s$^{-1}$) is not well matched by any of our models, although its column density is close to that we derive for Model 33, highlighting the danger of just relying on column densities comparison alone (e.g. as in \citet{Viti14}). 
The observed column density for methanol, $\sim$ 10$^{14}$ cm$^{-2}$, is well matched by several models, of which all but one have a gas density of 10$^5$ cm$^{-3}$; interestingly all of the shock models over-predict its column density. 

In summary, while a handful of single dish transitions are clearly not enough to characterize the multi-gas components of the nucleus of a galaxy, our ab initio method is able to certainly exclude ranges of densities, temperatures, and other physical parameters that do not affect the observed gas.
It is clear that single dish observations cannot be interpreted by assuming different species and transitions as unique tracers of a particular environment.  However this qualitative comparison already seems to indicate that while the CS (3-2) transition traces very dense gas at $\sim$ 100 K, it is likely that the CS (5-4) traces a hotter component and the HCO$^+$ traces a hot, possibly young, gas, at a lower density. Comparisons with interferometric observations will offer stringent tests to our method, as outlined by the example below. 

\subsection{NGC~253}

Again, as for the case of M82, here we simply perform an exercise to see how a priori modelling can help constrain some of the characteristics of the dense gas in another commonly studied galaxy.
\citet{Meier2015} spatially resolved several molecular species in the central kiloparsec 
of NGC~253 (at a distance of $\sim$ 350 pc) with ALMA at a resolution of $\sim$ 50 pc. They identified 27 molecular lines in 10 individual locations. \citet{Meier2015} performed a detailed analysis highlighting the rich chemistry in the central zone of this galaxy. They reported the integrated line intensities, derived column densities, and LVG analyses. Their Figure 5 shows an HCN/HCO$^+$ ratio as a function of position going from $\sim$ 0.8 to 1.6. Looking at Table 9 we find that this range is hard to match with any model and in fact those that are close to their best fit LVG model for the inner nuclear disk (positions 5--7), i.e at a gas density of n$_H$ =10$^{4.5-5.0}$ cm$^{-3}$ and a gas temperature of T = 60-120 K, all lead to a much higher HCN(1-0)/HCO$^+${1-0} ratio. This is consistent with \citet{Meier2015} conclusions that the main isotopologue HCN/HCO$^+$ line ratios
cannot alone be used as a unique diagnostic for starburst or AGN dominated galaxies, in this case probably owing to their high opacities. On the other hand, if we look at the 
individual theoretical intensities from Tables 7 and 8, we find that a good match for the HCN(1-0) transition can be obtained with Models 46 and 61 (10$^5$-10$^6$ cm$^{-3}$, 50-200K) but only at early times, i.e the gas does not seem to be in equilibrium. Interestingly, the gas traced by this transition does not seem to be affected by enhanced cosmic ray ionization rates. The HCO$^+$ (1-0) transition in the same nuclear region can be matched by Models 27, 39, and 54, differing in physical characteristics but all sharing the commonality that the cosmic ray ionization rate is higher than standard; more importantly the gas is at equilibrium, implying that the gas that HCO$^+$ is tracing is not the same as that traced by HCN(1-0), which is another possible reason why their ratio is a poor indicator in this occasion.
   
\citet{Meier2015} also observes typical shock tracers: SiO and HNCO. Making the assumption that SiO traces shocks they interpret the SiO/HCN line ratio as an indicator of gas that is experiencing shocks. Yet, their SiO abundance, $\sim$ 10$^{-9}$, which depending on the visual extinction gives
column densities of 2--8$\times$10$^{13}$ cm$^{-2}$, is not matched by our shock models but rather by models that have an enhanced cosmic ray ionization rate (Models 33, 36, 39, see Table 10). On the other hand, HNCO can be matched by several models, including a shock model (Model 67).  Hence, in this particular case, it seems that SiO can be used as a tracer of enhanced cosmic ray ionization gas. 

\par
Speculating further on the physical characteristics of each galaxy is beyond the scope of this paper  but it is clear that a comparison of several transitions and species with large grids of chemical models may lead to a more complete picture than simply comparing molecular line ratios, especially if both column densities and individual line intensities, with appropriate corrections based on the filling factor of each observation, are taken into consideration.

%

\section{Conclusions}

We present a grid of chemical models and provide the abundances, abundance ratios, and column densities  of the some of the most common tracers for a large parameter space in gas density, temperature, cosmic ray ionization rate, and radiation field (Tables 1--6). The theoretical abundances are then used as inputs to a radiative transfer code  to derive the theoretical intensities for the parameter space investigated by the chemical models (Tables 7, 8, 10--16). We finally provide examples of comparisons of observations from two galaxies with our theoretical line intensities. Our main conclusions, which are only valid within the parameter space of chemical models that we ran here, are as follows: (i) Time dependent effects on chemical abundances are only important for models in which the cosmic ray ionization rate is close to the galactic ionization rate. However the most affected species are, in fact, those most used to trace dense gas (HCN and CS). (ii) Line intensities ratios, such as HCN(1-0)/HCO$^+$(1-0), are not unique across the grid, leading to clear degeneracies in any RADEX solution. (iii) Individual species can be unique tracers of a particular energetic process, but only within specific physical conditions. (iv) A brief analysis of molecular observations from two  galaxies have shown how only relying on molecular line ratios can lead to erroneous conclusions on the physical conditions of the emitting gas and how therefore individual intensities, as well as column densities or fractional abundances, need to be taken into consideration.

\begin{acknowledgements}
The author acknowledges the AAO's Distinguished Visitor Scheme for their financial support during her visit at the AAO. The author thanks Dr. T. Greve and Dr. I. Jimenez-Serra for invaluable comments on the manuscript, and the anonymous referee for constructive comments that  improved the original version of the paper.

\end{acknowledgements}
\bibliographystyle{aa}
\bibliography{references}
\end{document}